# Four-Dimensional Usability Investigation of Image CAPTCHA

Junnan Yu*, Xuna Ma, Ting Han**

*School of Media & Design, Shanghai Jiao Tong University, Shanghai 200240, China*

*Junius@sjtu.edu.cn, **Hanting@sjtu.edu.cn

**Abstract:** Image CAPTCHA, aiming at effectively distinguishing human users from malicious script attacks, has been an important mechanism to protect online systems from spams and abuses. Despite the increasing interests in developing and deploying image CAPTCHAs, the usability aspect of those CAPTCHAs has hardly been explored systematically. In this paper, the universal design factors of image CAPTCHAs, such as image layouts, quantities, sizes, tilting angles and colors were experimentally evaluated through the following four dimensions: eye-tracking, efficiency, effectiveness and satisfaction. The cognitive processes revealed by eye-tracking indicate that the distribution of eye gaze is equally assigned to each candidate image and irrelevant to the variation of image contents. In addition, the gazing plot suggests that more than 70% of the participants inspected CAPTCHA images row-by-row, which is more efficient than scanning randomly. Those four-dimensional evaluations essentially suggest that square and horizontal rectangle are the preferred layout; image quantities may not exceed 16 while the image color is insignificant. Meanwhile, the image size and tilting angle are suggested to be larger than 55 pixels $\times$ 55 pixels and within $\pm 45°$, respectively. Basing on those usability experiment results, we proposed a design guideline that is expected to be useful for developing more usable image CAPTCHAs.

**Key Words:** Image CAPTCHAs, Usability, Eye tracking, Cognitive Model

# 1. Introduction

To distinguish human users from malicious script attacks, CAPTCHA (Completely Automated Public Turing test to tell Computers and Humans Apart) is widely deployed in human-machine systems, such as email services, online banking, social media applications etc. A well-designed CAPTCHA challenge should be easy for human users to solve while hard for bot to crack. Among the various CAPTCHA mechanisms proposed over the past 15 years, Text and Image CAPTCHAs are the dominate designs generally employed in nowadays [1, 2]. A text CAPTCHA usually includes several alphanumeric characters and human users are required to correctly recognize and input those characters to pass the verification [1]. However, to at effectively defend malicious attacks based on OCR (Optical Characters Recognition), those alphanumeric characters are usually distorted and added with background noises [3, 4]. Therefore, despite its wide deployment, Text CAPTCHA is essentially uncomfortable in visualization and hard to recognize [5]. Moreover, in the era of mobile computing, Text CAPTCHAs are less user-friendly on mobile devices which have smaller screens than PCs [6]. Text CAPTCHAs are also language dependent. It may be slower and less accurate for non-native speakers to solve English CAPTCHAs [7]. Recently, there is an increasing interest in deploying CAPTCHAs based on images. To pass an Image CAPTCHA challenge, a human user is required to follow certain instructions and make selections among several candidate images [2]. The benefits of Image CAPTCHAs lay in several folds [8]: Firstly, due to its nature of intuitiveness, Image CAPTCHA can be easily understood and solved by users with diverse cultural backgrounds. Secondly, in the era of mobile computing, it's a preferable option for mobile devices because of its simple verification procedure that involves only several finger swipes. Thirdly, Image CAPTCHAs are usually resistant to attacks and also visually comfortable. It is still a challenge for computer algorithms to effectively interpret the candidate images and make correct selections following CAPTCHA instructions. Therefore, Image CAPTCHAs are visually comfortable because no extra features, such as distortions or background noises, are required to impose on candidate images.

Due to the privileges provide by Image CAPTCHAs, a number of Image CAPTCHA mechanisms have been proposed. For instance, Asirra requires users to identify cats out of 12 candidate images ( arranged as 2 rows by 6 columns ) which include either cats or dogs [9]; reCAPTCHA presents users with random instructions to make selections among a set of 9 images that are arranged as 3 rows by 3 columns[10]. However, among all those Image CPATCHA designs proposed, the design factors that may affect the usability of CAPTCHAs have hardly been explored systematically.

In this study, we analyzed Image CAPTCHAs designs in nowadays, identified a set of universal design factors, such as image layout, quantity, size, tilting angle and color, and systematically evaluated the effects of those factors on the usability of Image CAPTCHA. Particularly, in addition to the three criteria: effectiveness, efficiency and satisfaction that are generally used for usability studies [11], we proposed a four-dimensional method for the usability evaluation of Image CAPTCHAs, which add a fourth dimension, eye-tracking. Because Image CAPTCHAs involves with various candidate images and layouts, the information collected by eye-tracking technology, such as scan path, heat map, etc., contribute an extra dimension to understand users' cognitive behaviors of solving Image CAPTCHAs. Such a four-dimensional usability investigation provided panoramic information on the usability of image CAPTCHAs. The usability experiment results presented in this paper may serve as a design guideline that is expected to be helpful for developing usable image CAPTCHAs.

The remainder of this paper is organized as follows. In section 2, we present the related works. Section 3 detailed the experimental designs. Section 4 is focused on the experimental results and discussions. Section 5 and 6 are dedicated to conclusion and limitations, respectively.

## 2. Related Work

A number of Image CATPCHA designs have been reported previously. Elson proposed Asirra, which requires users to identify cats from a set of 12 candidate images that includes either cats or dogs [9]. As shown in Fig. 1(a), the candidate images of Asirra were arranged as a 2 rows by 6 columns layout. The daily-updated image database from Petfinder.com were used, which includes millions of photographs that provide reasonable security against automatic scripts and bots. More than 300 participants were recruited to test the usability of Asirra. It is revealed that, because all participants were familiar with both cats and dogs, about 99.6% of the CAPTCHAs were correctly solved within 30 seconds. Confident CAPTCHA [13] and KittenAuth [14] are similar with reCAPTCHA: both of them employ 9 candidate images and users are required to pick designated image(s) according to certain instructions. The verification process of image CAPTCHAs proposed by Chew [15] requires users to summarize and distinguish the common topics shared by all the images presented in a CAPTCHA. Moreover, IMAGINATION [17] and SEMAGE [18] also involved the identifying of common characters shared by the candidate images. For Click-based Graphical CAPTCHAs [19], the verification process requires users to click candidate images with a particular sequence following the instruction. In addition, several schemes based on face recognitions among the candidate images are also proposed [20-24]. Usability comparison has been reported for four of the Image CAPTCHAs mentioned above: Asirra, ESP-PIX, SQ-PIX, and IMAGINATION, which revealed that Asirra and ESP-PIX are more preferable by users [25].

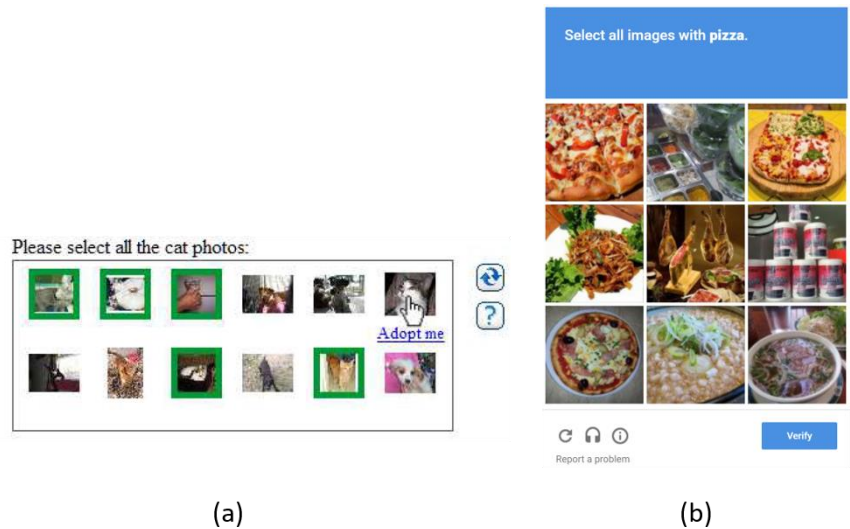

**Fig. 1**. Example of two Image CAPTCHA designs: (a) Asirra; (b) Google reCAPTCHA

## 3. Method

The usability of each design factor was evaluated through four variables: eye tracking, efficiency, effectiveness and satisfaction. The eye activities during the solving of CAPTCHAs were recorded through an eye tracker. The efficiency was defined as the averaged time for participants to solve a particular type of CAPTCHA while the effectiveness was the accuracy rate of solving that kind of CAPTCHA. The satisfaction was obtained through questionnaires and face-to-face interviews with the participants.

### 3.1 Participants

We recruited 37 participants (17 males and 20 females) during the experiment sessions. The participants' ages ranged from 18 to 25, with an average value of 21.43 and a standard deviation of 2.17. All subjects were students from Shanghai Jiao Tong University and 21 of them were undergraduate students while the remaining were graduate students. All subjects understood the simple English instructions such as "Submit" appeared in the tests. Meanwhile, they were all experienced computer users and had encountered Image CAPTCHAs before participating this experiment. None of them was of light color blindness or color blindness. Also, they all had no trouble reading on a computer screen or solving Image CAPTCHAs with a mouse.

**3.2 Apparatus**

The experiment was performed in a lab environment. The eye tracker used in current study was Tobii T60, which was controlled by a laptop running Windows 8.1. The data acquisition rate of the eye tracker was 60 Hz and the angular resolution of the eye tracker is 0.5 degree. Testing CAPTCHAs were displayed on a 19-inch screen set at a resolution of 1280 pixels ×1024 pixels. To obtain the maximum performance of the eye tracker, we used the following parameters that were recommended by the manufacturer: the distance between screen and participant was about 60 cm; the test CAPTCHA was displayed around the center of the screen; the height of the chair was adjustable so that the eyes of a participant were horizontally parallel with the middle of the screen. The eye tracking data was recorded by Tobii Studio Pro 4.0, an analyzing software provided by the manufacturer. Before the test session, each participant was instructed to finish a calibration session on Tobii Studio Pro 4.0 to guarantee the accuracy of eye tracking data. The recorded eye-tracking data include the trace of eye gazing on the screen, as well as how long the eyes focused on a particular area on the screen. The test CAPTCHAs were generated on a remote server and downloaded to the local browser, FireFox 5.6, in the form of webpages. Standard input devices, such as keyboard and mouse, were provided for the solving of test CAPTCHAs. The solving time and accuracy rate of each CAPTCHA design were stored on the remote server for further analysis.

## 3.3 Experimental Design

Basing on CAPTCHA schemes mentioned above, the design factors evaluated here included the layout of images, number of images, size of images, tilting angle of images and color of images. Because all those factor correlated with one another, for the evaluation of a certain factor, all other factors are set at a fixed value which is estimated the most usable or has been used by most previously reported image CAPTCHA designs. All images used in this study are obtained from the open access project Asirra [27]. To solve a test CAPTCHA, participants are instructed to select cat images from a set of candidate images that contain either cats or dogs. One of the test CAPTCHAs is shown in Figure 2 for demonstration purpose.

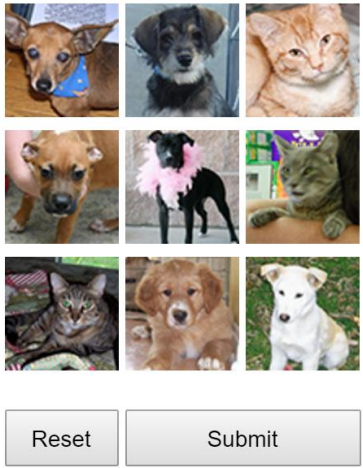

**Fig. 2.** An example of the test CAPTCHAs used in our experiments. The Image CAPTCHA employs a 3×3 layout and participants are instructed to select all images that contain cats and then submit their selection.

### 3.3.1 Image layout

As depicted in Figure 3, the six layouts evaluated in this study are Square (S), Horizontal Rectangle (HR), Vertical Rectangle (VR), Upright Triangle (UT), Inverted Triangle (IT), and Trapezoid (T). Each layout employs 16 images and the size of each image is 70 pixels ×70 pixels. All images are colored and the subject of each image is not rotated (set at 0°).

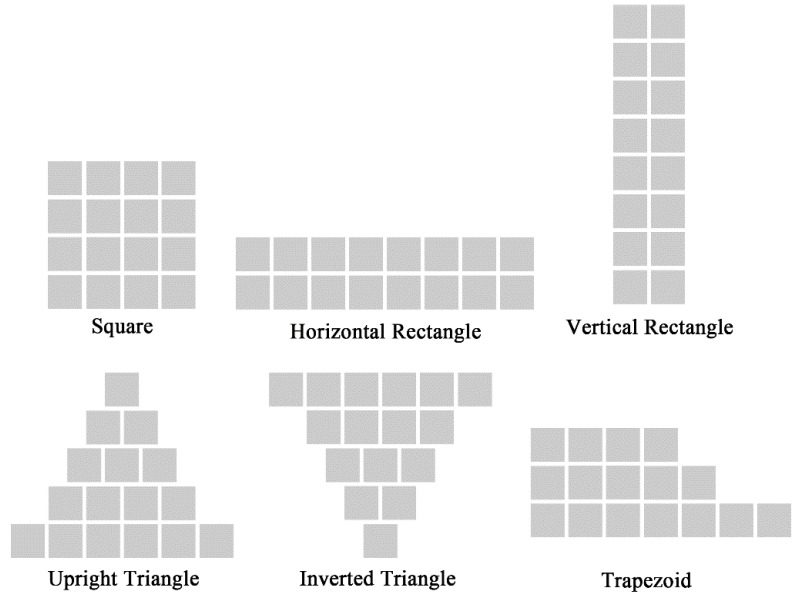

**Fig. 3**. The six layouts evaluated in this study: Square (S), 4 rows ×4 columns; Horizontal Rectangle (HR), 2 rows ×8 columns; Vertical Rectangle (VR), 8 rows ×2 columns; Upright Triangle (UT), 5 rows ×1~6 columns; Inverted Triangle (IT), 5 rows ×6~1 columns and Trapezoid (T), 3 rows ×4~7 columns. "▪" represents the position of each candidate image.

### 3.3.2 Image quantity

In this section, the numbers of candidate images in each CAPTCHA design are 4, 9, 16 and 25. Square layout is employed and the candidate images are arranged as 2 ×2, 3 ×3, 4 ×4 and 5 ×5 (rows ×columns). The size of each image is 70 pixels ×70 pixels. All images are colored and the subject on each image is not tilted (set at 0 °).

### 3.3.3 Image size

As shown in Fig. 4, images of five different sizes are evaluated: 25 pixels×25 pixels, 40 pixels×40 pixels, 55 pixels×55 pixels, 70 pixels×70 pixels, and 85 pixels×85 pixels. Except for the size of candidate images, all other design factors are kept the same: Square layout is employed which includes 9 images and arranges as 3 rows ×3 columns. All images are colored and the subject on each image is not tilted (set at 0 °).

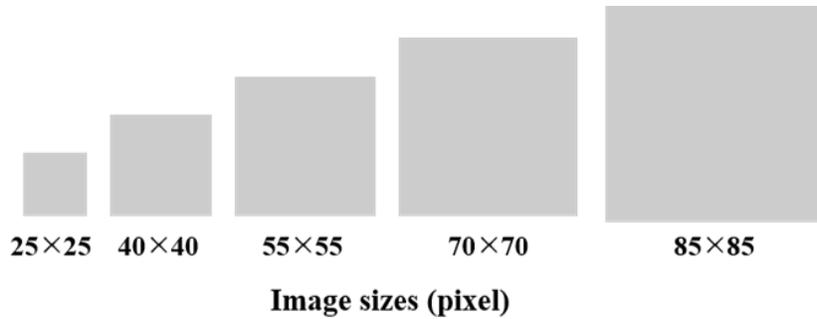

**Fig. 4**. The five testing image sizes: 25×25 pixels, 40×40 pixels, 55×55 pixels, 70×70 pixels, and 85×85 pixels. "▪" represents the candidate image.

**3.3.4 Tilting angle of images**

As depicted in Fig. 5, the subject on each candidate image is tilted by a certain angle. The 8 angles evaluated are 0°, 45°, 90°, 135°, 180°, 225°, 270°, and 315°. Among all those tilting angles, 0° is chosen as the reference for which the subject of an image is at an ordinary post. All other tilting angles are rotated clockwise with respect to the reference one. Except for the tilting angle, all other design factors are kept the same: Square layout is employed which includes 9 images and arranges as 3 rows ×3 columns. The size of each image is 70 pixels ×70 pixels. All images are colored.

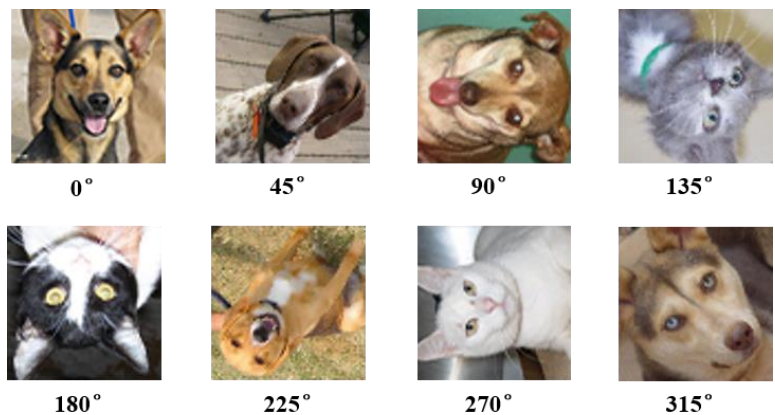

**Fig. 5**. Illustration of different tilting angles: 0°, 45°, 90°, 135°, 180°, 225°, 270°, and 315°.

### 3.3.5 Image color

In this section, the effect of colored or monochrome images are evaluated. Except for the color of candidate images, all other factors are kept the same: Square layout is employed which includes 9 images and arranges as 3 rows ×3 columns. The size of each image is 70 pixels ×70 pixels. The subject of each image is not tilted (set at 0 °).

### 3.3.6 Satisfactory questionnaire

For each design factor, the corresponding satisfaction questionnaire focuses on the following three aspects: Q1. It is visually comfortable; Q2. It's easy to be recognized and solved; Q3. It's appropriate for application. Participants are instructed to rate each design factor in terms of (Q1) visual comfort, (Q2) ease of use and (Q3) appropriateness for application. The scores are defined by a 5-point Likert-scale (1=strongly disagree, 5=strongly agree) and all statements were scored in a positive scale. The satisfactory result of each design factor is averaged among all the participants and presented in tables.

## 3.4 Procedure

The experiments were conducted with two consecutive stages: preparation and test. During the preparation stage, the testing apparatuses were reset and the experiment purpose and tasks were introduced to participants. All participants were also aware that the experiment was anonymous and the usage of any information collected was restricted for this study only. After that, each participant was instructed to get familiar with the experiment apparatuses such as the input device, the interface of the image CAPTCHAs. The eye tracker was also individually calibrated for each participant to make sure that the eye-tracking data were correctly collected. Before proceeding to the test stage, participants were encouraged to ask any questions regarding the experiment if there were any unclarified issues. During the test stage, the test CAPTCHAs were presented one by one and the participant was left alone in the lab without any disturbance. The solving time and user input of each CAPTCHA were indexed on a remote server for further analysis. The eye-tracking data were automatically collected and stored on the control computer. The first 3 CAPTHCAs presented in the test were aimed at helping the participant to get familiar with the experiments and the data of those CAPTCHAs were eliminated. After solving all the test CAPTCHAs, an online questionnaire was presented to acquire the participant's subjective opinions on each design factor. The overall test stage takes about 25 minutes, which is within the threshold that a participant can stay focus. During the test stage, an experiment instructor waited outside the lab in case the participant need any support. After finishing the experiments, each participant was paid with a small amount of cash to appreciate his/her cooperation.

## 4. Results and discussion

Five design factors are evaluated in our study and each design factor includes 2 to 8 variables. Therefore, there are 25 variables evaluated in this contribution. For each variable, five CAPTCHAs are presented to the participant. Given that 37 participants are recruited, the solving time, accuracy rate and eye-tracking data for each variable are averaged over $37 \times 5=185$ individual measurements. However, the eye-tracking data of two participants were not effectively collected and were therefore eliminated. SPSS 22.0 was used for data analysis. A repeated measures ANOVA was used with a 95% confidence interval to determine significant differences across different CAPTCHAs. Given significance, Post-hoc LSD comparisons were used to reveal the locus of significant differences. Friedman test was used to measure the significance of the user preference due to its nonparametric nature.

### 4.1 Cognitive processes revealed by eye-tracking

The data extracted from the eye tracker are presented by heat maps and gaze plots. The heat maps provide insights on how looking is distributed on the images of a CAPTHCA，which represents the focus of visual attention averaged among all the participants. The gaze plot, however, shows the location, order, and time spent looking at locations on the image CAPTCHAs, which reveals the time sequence of looking for each participant.

The heat maps of the six different layouts are presented in Fig. 6, which is obtained through accumulating the eye-gazing data of all participants. It is evident that the participant's attentions are almost evenly distributed within each layout. During the test processes, all candidate images used for the test CAPTCHAs are randomly displayed and their contents vary from one another: the cat or dog on each image is different in color, species, or pose; there are also fluctuations for the image background, light condition and so on. Therefore, the averaged attention distribution visualized by heat maps indicates that participants are nearly equally attracted by those various images displayed on each CAPTCHA. There is no preferred gazing location on any of the layouts tested in our experiments. Those heat maps suggest that, for the design of Image CAPTCHAs, the exposure of each image position on a layout is equal and there is no need to specially consider the visibility. It is also worthy to mention that, for images that are ambiguous (for example, a black cat that looks like both cat and dog), it may attract more attention because participants need extra time to correctly recognized the subject on an image. This may eventually cause "hot spots" on a heat map. However, such property is associated with the candidate images, not the layouts.

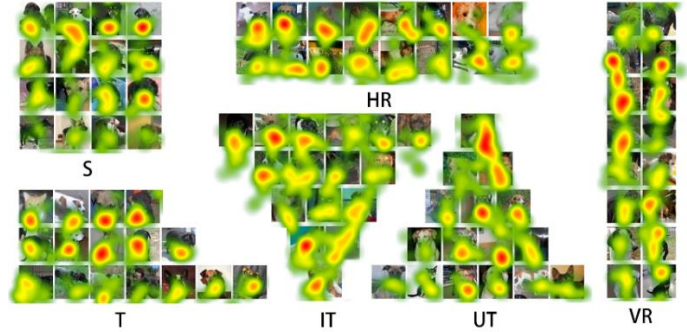

**Fig. 6**. Heat maps of the six layouts: Square (S), Horizontal Rectangle (HR), Vertical Rectangle (VR), Upright Triangle (UT), Inverted Triangle (IT), and Trapezoid (T).

Besides the distribution of attention, it is also important to know how the eyes are moved from one image to another. As shown in Figure 7, the gaze plots of all the six different layouts can be classified into three types: S type, Z type and Random type. For the S type, the eye gazing directions are reversed for two consecutive rows. For example, if the eyes move from left to right till the end of one row, then the eyes scan from the right side of the next row till the leftmost image of that row. For the Z type, the eye gazing directions are the same for all rows: either from left to right or from right to left. However, for Random type, the eye gazing is unpredictable and randomly goes through all the candidate images. It should be noted that, for some gaze plots, the movements of eyes may not exactly follow the criteria of S or Z type. Such cases will be attributed to the closest type that best fit them. For example, the eyes may randomly go back and forth on three closely arranged images while the eye gazing on the rest of other images still follow the characters of S type. Such example will eventually be categorized as S type. In addition, it is interesting that all participants gazing from top to bottom, regardless of the different layouts. This reflects the daily reading habits. Moreover, for each individual participant, he/she will stay with the same eye gazing type to finish all the test CAPTCHAs. This agree with the theory of stereotypical scanning strategies that a viewer will use the same sequence of looking for different test objects [26]. The gazing plots also indicate that, participants can only focus on one image at a time, regardless of the size of images and how many images are displayed.

The percentage of each gazing type is listed in Table 1. Because the Vertical Rectangle layout includes only two columns, the boundary between S and Z type is blurred. Therefore, we count S and Z type together for Vertical Rectangle layout in the table. It is clear that, most participants inspected the images of a CAPTCHA following a certain order and usually less than 30% of the participants graze randomly. However, for the Trapezoid layout, the percentage of random gazing is 42.86%. Given that the Trapezoid layout is irregular comparing with the other five layouts, this high percentage of random gazing implies that irregular layouts may cause confusion to participants and they therefore tended to randomly search CAPTCHA images. During the face-to-face interviews, participants also mentioned the inconvenience of irregular layouts. The comparison within the five relatively regular layouts indicates that, square and rectangle layouts induce less random gazing than triangle layouts. Preference on square and rectangle layout are also found in the face-to-face interviews and online questionnaires after the test. It is also interesting that more than 86.67% of the participants will double-check their selections before submission.

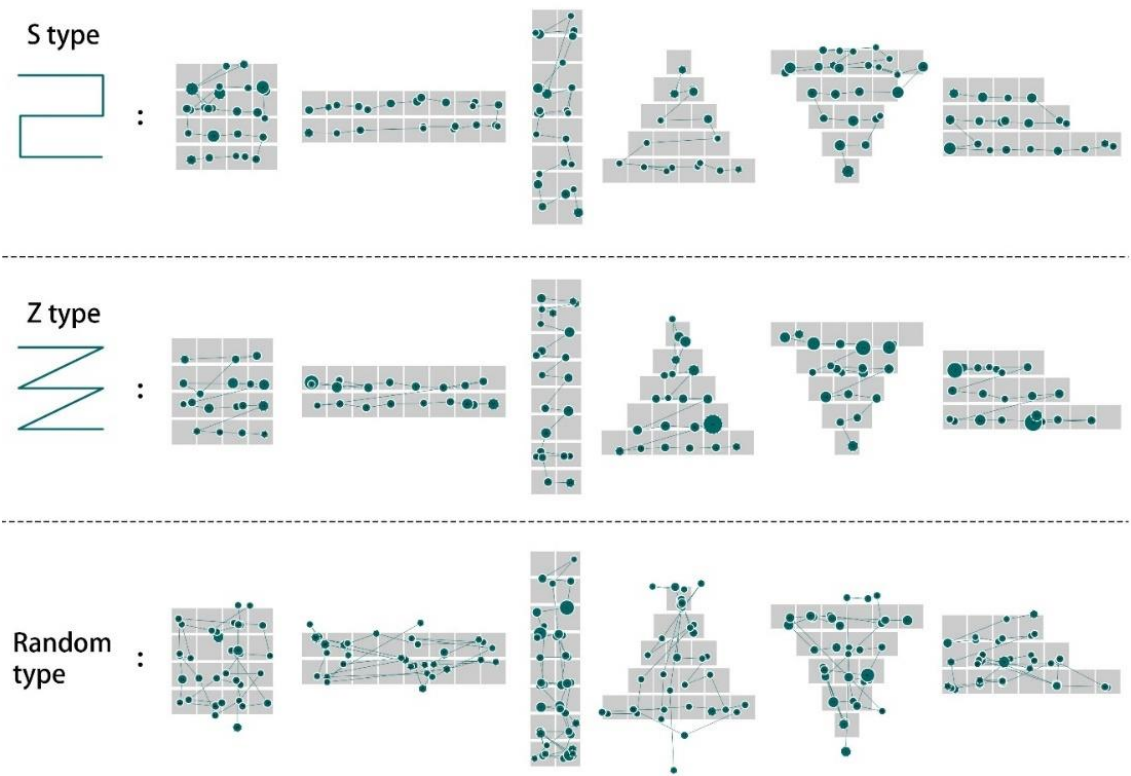

**Fig. 7**. Extracted pattern samples of participants' scan paths: S type (the paths were similar with the mirror-reversed S pattern), Z type (the scan paths followed zigzag pattern), and Random type (the scan paths were irregular).

**Table 1**. Participants' scan-path-pattern distributions of the testing layouts

|  | S type (%) | Z type (%) | Random (%) |
|---|---|---|---|
| **Square** | 75.00% | 14.29% | 10.71% |
| **Horizontal Rectangle** | 64.29% | 21.43% | 14.29% |
| **Vertical Rectangle** |  | 82.14% | 17.86% |
| **Upright Triangle** | 42.86% | 28.57% | 28.57% |
| **Inverted Triangle** | 46.42% | 28.57% | 25.00% |
| **Trapezoid** | 39.29% | 17.86% | 42.86% |

The time spent on looking at CAPTCHA images with different layouts are listed in Table 2. For all the six different layouts, the eye gazing time are similar. Considering that each layout includes 16 images, the averaged time spent on looking at a single image is about 0.5 seconds. We also compared the eye gazing time for Square layout that employ different quantity of images, which is shown in Table 3. It is interesting that, although the total gazing time increases with the number of candidate images, the averaged gazing time on a single image actually decreased. This is probably because that, with the increasing of candidate images, participants are less patient and therefore spent less time on each candidate image. Those results indicate that, 9~16 candidate images are a better balance between user's patience and enough candidate images to provide better security.

Table 2. Participants' average duration time of images of the testing layouts

|  | Square | | Horizontal Rectangle | | Vertical Rectangle | | Upright Triangle | | Inverted Triangle | | Trapezoid | |
| --- | --- | --- | --- | --- | --- | --- | --- | --- | --- | --- | --- | --- |
|  | AVG | SD | AVG | SD | AVG | SD | AVG | SD | AVG | SD | AVG | SD |
| **Duration of each layout (s)** | 8.99 | 1.95 | 8.97 | 1.95 | 9.09 | 3.01 | 8.71 | 2.59 | 8.44 | 2.55 | 8.43 | 2.21 |
| **Duration of each image (s)** | 0.56 | -- | 0.56 | -- | 0.57 | -- | 0.55 | -- | 0.53 | -- | 0.53 | -- |

Table 3. Participants' average duration time of images of different numbers

|  | 4 pieces | | 9 pieces | | 16 pieces | | 25 pieces | |
| --- | --- | --- | --- | --- | --- | --- | --- | --- |
|  | AVG | SD | AVG | SD | AVG | SD | AVG | SD |
| **Duration of each number (s)** | 2.74 | 0.71 | 4.96 | 1.94 | 8.99 | 1.95 | 10.03 | 2.66 |
| **Duration of each image (s)** | 0.69 | -- | 0.55 | -- | 0.56 | -- | 0.40 | -- |

According to those eye-tracking data and the EPIC cognitive architecture [27], we summarize a generalized cognitive model for the solving of Image CAPTCHAs, which is shown in Fig. 8. The cognitive procedure proposed here consists of three steps: (i) decoding of instructions; (ii) searching and selecting images; (iii) submitting results. In the first step, CAPTCHA instructions is extracted by eyes and decoded in cognitive processor, which generates commands for eye movements. In the second step, those commands are delivered to the Ocular Motor Processor, which triggers the eyes to scan test images. A perceived image's content, property, and position are delivered to the Cognitive Processor, which determines if that candidate image match the instructions decoded in the first step. For images that match the instructions, commands are sent to Manual Motor Processor to make selections by means of mouse/touchpad click. For images that do not match the instruction, Ocular Motor Processor is enabled to move the eyes to focus on a new image. After all candidate images of a CAPTCHA are inspected, it comes the third step: Submission. The eyes are driven to the submission button and Cognitive Processor sends commands to Manual Motor Processor, which triggers hand movements to click the submission button and finish the whole cognitive cycle of solving an image CAPTCHA.

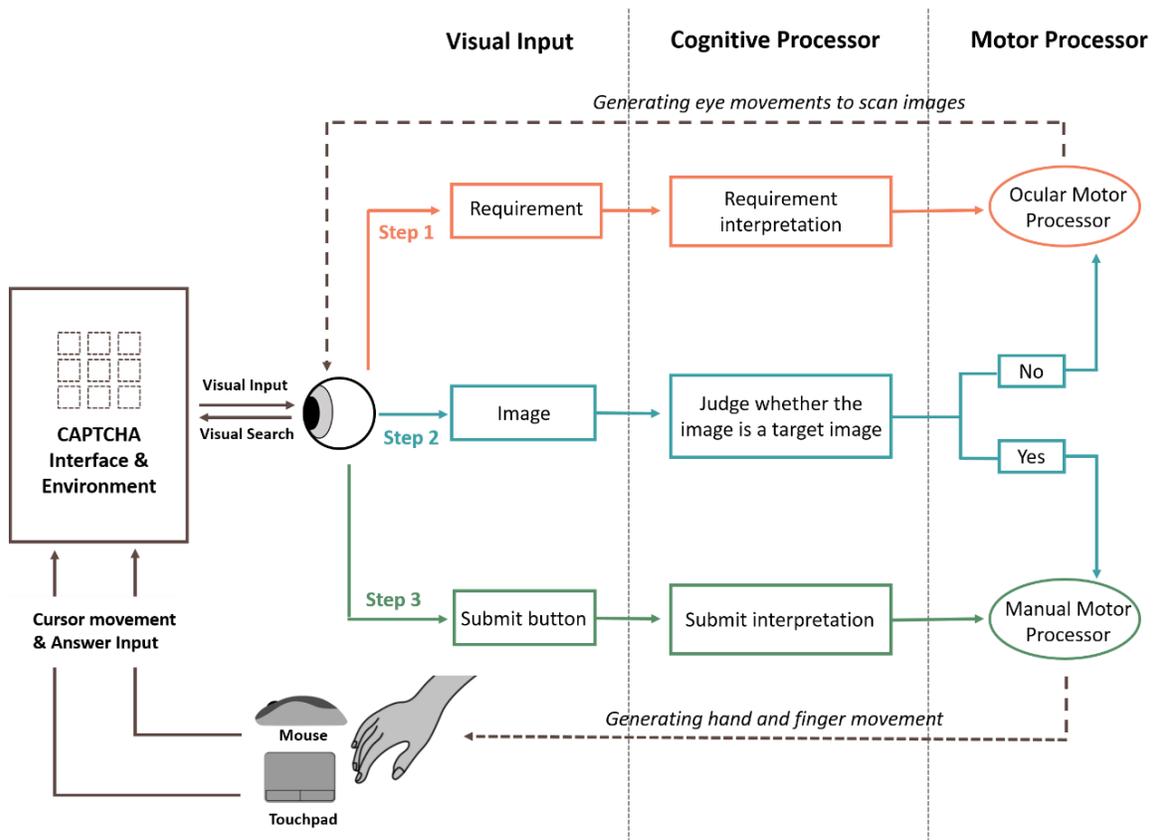

**Fig. 8**. Cognitive model of solving an image CAPTCHA. Solid arrows represent information flow paths, dashed arrows denotes mechanical control or connections. The processors run independently and in parallel both with each other and CAPTCHA environment module.

## 4.2 Efficiency, Effectiveness and Satisfaction Evaluation

### 4.2.1 Image layout

The average solving time (M) and its corresponding standard deviation (SD) of the six different image layouts are depicted in Fig. 9: Square (M=9.57s, SD=4.02s), Horizontal Rectangle (M=10.08s, SD=3.73s), Vertical Rectangle (M=10.10s, SD=4.71s), Upright Triangle (M=10.03s, SD=4.49s), Inverted Triangle (M=9.68s, SD=3.86s) and Trapezoid (M=9.31s, SD=3.97s). A repeated measures ANOVA with a Greenhouse-Geisser correction indicated that those solving times have statistical difference [$F(4.58, 842.53)=1.90$, $p=0.039$]. The accuracy rate is also given in Fig. 9 for the six image layouts: Square (91.35%), Horizontal Rectangle (89.73%), Vertical Rectangle (92.43%), Upright Triangle (89.19%), Inverted Triangle (91.89%) and Trapezoid (83.78%). The satisfaction questionnaires are summarized in Table 4 in terms of (Q1) visual comfort, (Q2) ease of use and (Q3) appropriateness for application. Friedman tests return that the three dependent variables are all statistically significant: visual comfort [Q1, $\chi2(2) =112.16$, $p<0.001$], ease of use [Q2, $\chi2(2) =84.451$, $p<0.001$], and appropriateness for application [Q3, $\chi2(2) =114.32$, $p<0.001$].

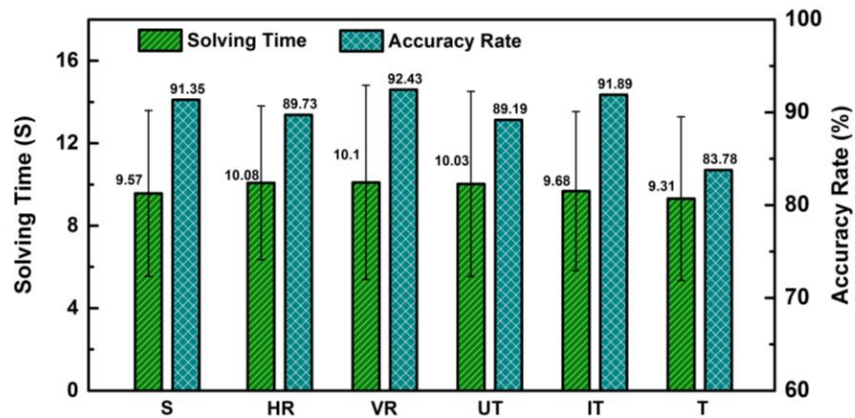

**Fig. 9**. The average solving time and accuracy rate of the six layouts: Square (S), Horizontal Rectangle (HR), Vertical Rectangle (VR), Upright Triangle (UT), Inverted Triangle (IT), and Trapezoid (T).

Table 4. Satisfaction of CAPTCHAs based on different layouts.

| | Square | | Horizontal Rectangle | | Vertical Rectangle | | Upright Triangle | | Inverted Triangle | | Trapezoid | |
|---|---|---|---|---|---|---|---|---|---|---|---|---|
| | AVG (1-5) | SD | AVG (1-5) | SD | AVG (1-5) | SD | AVG (1-5) | SD | AVG (1-5) | SD | AVG (1-5) | SD |
| **Q1.** It's visually comfortable | 3.97 | 0.96 | 3.37 | 0.85 | 2.37 | 1.03 | 3.13 | 1.28 | 2.57 | 1.14 | 2.57 | 0.97 |
| **Q2.** It's easy and efficient to recognize the images | 3.60 | 0.93 | 3.40 | 0.97 | 3.17 | 1.15 | 3.17 | 1.15 | 2.77 | 1.10 | 2.63 | 1.03 |
| **Q3.** It's appropriate for wide application | 3.77 | 0.94 | 3.30 | 0.79 | 2.50 | 1.01 | 2.93 | 1.20 | 2.43 | 0.94 | 2.43 | 0.94 |

ANOVA shows the solving time of image CATPCHAs with different layouts are similar, however, no definite conclusions can be drawn about the effects of layout on solving time. The accuracy rate shows that Trapezoid is the lowest one, which is about 6%~9% lower than any other five layouts. This may be attributed to its irregular layout. As discussed in the eye tracking session, about 42% of the participants randomly checked the candidate images displayed on Trapezoid layout, which is 10%~20% higher than the other five layouts. Due to such chaotic scan paths, participants may miss some candidate images, which decreased its accuracy rate. Despite Trapezoid layout, the accuracy rates of the other five layouts are similar with a variation that is less than 3%. Satisfaction questionnaires indicate that, Square and Horizontal Rectangle are the most preferred layouts in terms of visual comfort, easy to use and appropriate for application.

### 4.2.2 Image quantity

The averaged solving time (M) and its corresponding standard deviation (SD) are depicted in Fig. 10 for CAPTCHAs using different quantities of images: 4 images (M=3.86s, SD=1.74s), 9 images (M=5.86s, SD=2.29s), 16 images (M=9.57s, SD=4.02s) and 25 images (M=12.76s, SD=4.48s). A repeated measures ANOVA with a Greenhouse-Geisser correction indicated that those solving times are statistically different [$F(2.33, 428.46)=333.357$, $p<0.001$]. Fig. 10 also shows the accuracy rates for different number of images: 4 images (98.92%), 9 images (92.30%), 16 images (91.35%) and 25 images (83.78%). The satisfaction questionnaires regarding different image sizes are summarized in Table 5 in terms of (Q1) Visual Comfort, (Q2) Ease of Use and (Q3) Appropriateness for Application. Friedman tests return that the three dependent variables are all statistically significant: Visual Comfort [Q1, $\chi 2(2)$ =83.85, $p<0.001$], Ease of Use [Q2, $\chi 2(2)$ =78.66, $p<0.001$], and Appropriateness for Application [$\chi 2(2)$ =81.94, $p<0.001$].

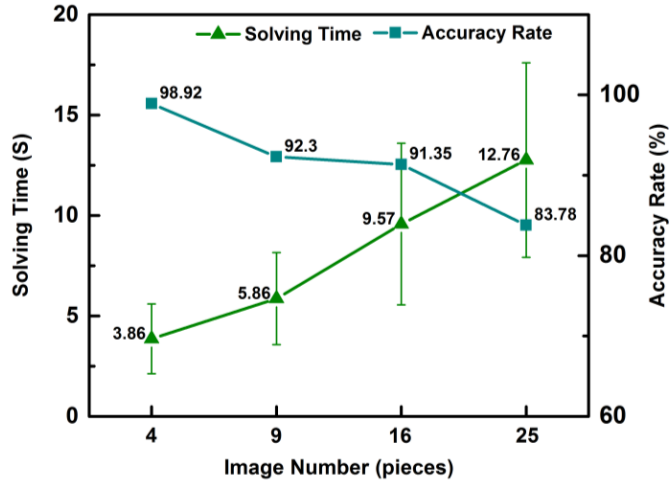

**Fig. 10**. Averaged solving time and accuracy rate of CAPTCHAs based on different image numbers: 4 pieces, 9 pieces, 16 pieces, and 25 pieces.

**Table 5**. Satisfaction of CAPTCHAs based on different image numbers.

|  | 4 pieces | | 9 pieces | | 16 pieces | | 25 pieces | |
| --- | --- | --- | --- | --- | --- | --- | --- | --- |
|  | AVG (0-5) | SD | AVG (0-5) | SD | AVG (0-5) | SD | AVG (0-5) | SD |
| **Q1.** It's visually comfortable | 4.37 | 0.89 | 4.23 | 0.73 | 3.23 | 0.90 | 1.97 | 0.93 |
| **Q2.** It's easy and efficient to recognize the images | 4.53 | 0.86 | 4.13 | 0.68 | 3.03 | 0.96 | 1.97 | 1.13 |
| **Q3.** It's appropriate for wide application | 4.03 | 1.03 | 4.37 | 0.61 | 2.97 | 1.01 | 1.86 | 1.01 |

The solving time increases linearly with respect to the number of images presented on a CAPTCHA, which is because of the higher workload to recognize more images. Those solving time are further compared with the eye-gazing times that were recorded by the eye tracker and displayed in Table 3. It is interesting to note that the eye-gazing time accounts for 70%~90% of the time that are used for solving CAPTCHAs, which indicates that the solving time of Image CAPTCHAs is mainly due to the recognition of candidate images. Subtracting the eye-gazing time on images from CAPTCHA solving time reveals that the time spent on making selections and submitting is about 1~2.5 seconds. The highest accuracy rate for 4 images, 98.92%, is because with fewer images, the probability of making mistakes is also lower. Meanwhile, participants may be more concentrated for just 4 images. On the contrary, with 25 images, the lowest accuracy rate of 83.78% is observed. This can be explained that, with increased candidate images, it is more likely to make wrong selections. Meanwhile, participants were also less patient and remain focused to recognize 25 images. This agrees with the previous eye-tracking data that, the eye-gazing time a participant spend on each image for 25 images is 42% less than that of 4 images. However, for 9 or 16 candidate images, the accuracy rates are similar, indicating that they are equally within the participant's tolerance. Satisfaction questionnaires further indicated that participants expect fewer images on a CAPTCHA. Although the scores of both 4 and 9 images are more than 4 among all three variables, it is worthy to mention that CAPTCHAs with fewer candidate images would be less secure against bots. Therefore, a quantity of 9 images is the most recommended number of candidate images in terms of solving time, accuracy rate, satisfaction and security.

### 4.2.3 Image Size

The averaged solving time (M) and its corresponding standard deviation (SD) are depicted in Fig. 11 for the five image sizes that have been evaluated: 25pixels $\times$ 25pixels (M=10.07s, SD=5.26s), 40pixels $\times$ 40pixels (M=6.92s, SD=2.87s), 55pixels $\times$ 55pixels (M=5.90s, SD=2.53s), 70pixels $\times$ 70pixels (M=5.86s, SD=2.29s) and 85pixels $\times$ 85pixels (M=5.73s, SD=2.29s). A repeated measures ANOVA with a Greenhouse-Geisser correction indicated that those solving times are statistically different [$F(2.76, 508.20)=75.52$, $p< 0.001$]. Fig. 11 also shows the accuracy rate of the five image sizes: 25pixels $\times$ 25pixels (58.92%), 40pixels $\times$ 40pixels (89.19%), 55pixels $\times$ 55pixels (89.19%), 70pixels $\times$ 70pixels (92.30%) and 85pixels $\times$ 85pixels (90.81%). The satisfaction questionnaires regarding different image sizes are summarized in Table 6 in terms of (Q1) Visual Comfort, (Q2) Ease of Use and (Q3) Appropriateness for Application. Friedman tests return that the three dependent variables are all statistically significant: Visual Comfort [Q1, $\chi2(2)$ =110.89, $p<0.001$], Ease of Use [Q2, $\chi2(2)$ =108.67, $p<0.001$], and Appropriateness for Application [Q3, $\chi2(2)$ =108.71, $p<0.001$].

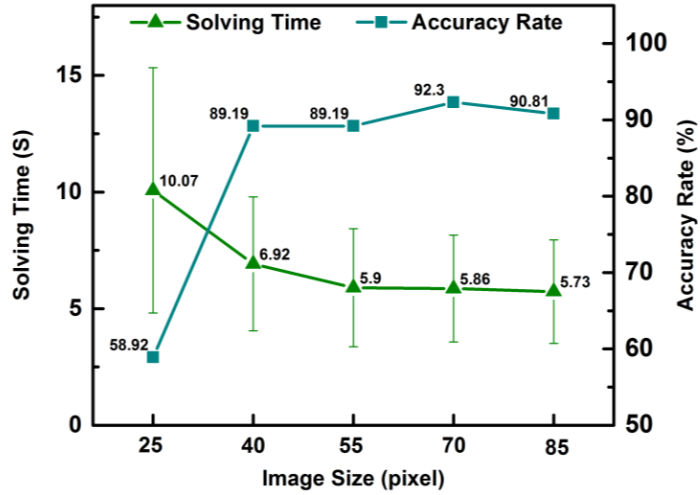

**Fig. 11**. The averaged solving time and accuracy rate of CAPTCHAs based on different image sizes: 25 pixels ×25 pixels, 40 pixels ×40 pixels, 55 pixels ×55 pixels, 70 pixels ×70 pixels and 85 pixels ×85 pixels.

**Table 6**. Satisfaction of CAPTCHAs based on different image sizes.

|  | 25×25 pixels | | 40×40 pixels | | 55×55 pixels | | 70×70 pixels | | 85×85 pixels | |
| --- | --- | --- | --- | --- | --- | --- | --- | --- | --- | --- |
|  | AVG (0-5) | SD | AVG (0-5) | SD | AVG (0-5) | SD | AVG (0-5) | SD | AVG (0-5) | SD |
| **Q1.** It's visually comfortable | 1.27 | 0.78 | 2.13 | 0.57 | 2.87 | 0.90 | 3.90 | 0.96 | 4.43 | 0.77 |
| **Q2.** It's easy and efficient to recognize the images | 1.20 | 0.76 | 2.07 | 0.83 | 2.83 | 0.87 | 3.80 | 0.89 | 4.53 | 0.86 |
| **Q3.** It's appropriate for wide application | 1.13 | 0.73 | 1.97 | 0.81 | 2.67 | 0.99 | 3.87 | 1.01 | 4.27 | 0.91 |

The averaged solving time dropped from 10.07s to 5.9s when the image sized increased from 25 pixels ×25 pixels to 55 pixels ×55 pixels. However, for image sizes that are larger than 55 pixels ×55 pixels, the solving time remain unchanged around 5.8 seconds. The much longer solving time for smaller images indicated that it took more time for participants to recognize the subjects on those images. This is because the subjects on smaller images are usually much more blurred comparing with those on the large images. However, for images that are large enough, more than 55 pixels × 55 pixels for example, the subject on an image is clear enough to be recognized. Therefore, it takes essentially the similar efforts to recognize subjects on such images and the solving time does not change much with the further increasing of image size. Accompany with the longer solving time for small images like 25 pixels × 25 pixels, the accuracy rate is also lower, which is again because that the subjects on those images are hard to recognized. Satisfaction questionnaires indicated that participants generally believed that larger images are better than smaller ones in terms of visual comfort, easy to solve and preferable in actually application. Given that the rating of 70 pixels × 70 pixels in all respects are above 3.8 out of a 5-point scale, it is concluded that images that are larger than such a size will be sufficient for user satisfaction. Those results also agree with Elson's findings [9] that smaller images degraded both solving time and accuracy. Although users may believe that the larger an image, the better, the experimental results present in this study reveal that the increasing of image size do not further improve the solving time and accuracy rate. In addition, the workload of processing large images are also higher on the remote servers, which may also bring inconveniences to those who slow Internet connections [9, 28].

### 4.2.4 Tilting angle of images

The averaged solving time (M) and its corresponding standard deviation (SD) is depicted in Fig. 12 for CAPTCHA images with different tilting angles: 0°(M=5.86s, SD=2.29s), 45°(M=6.33s, SD=2.24s), 90°(M=5.83s, SD=1.88s), 135°(M=7.24s, SD=3.21s), 180°(M=7.03s, SD=2.48s), 225°(M=7.09s, SD=3.83s), 270°(M=6.01s, SD=2.47s) and 315°(M=5.98s, SD=2.49s). A repeated measures ANOVA with a Greenhouse-Geisser correction indicated that those solving times are statistically different [$F(5.97, 1021.49)=11.73, p<0.001$]. The accuracy rate of each tilting angle is also given in Fig. 12: 0°(92.30%), 45°(90.27%), 90°(92.22%), 135°(89.19%),

180 °(88.65%), 225 °(88.11%), 270 °(92.43%), and 315 °(91.89%). The satisfaction questionnaires regarding different tilting angles are summarized in Table 7 in terms of (Q1) Visual Comfort, (Q2) Ease of Use and (Q3) Appropriateness for Application. Friedman tests return that the three dependent variables are all statistically significant: Visual Comfort [Q1, Q1, $\chi2(2)$ =176.06 $p<0.001$], Ease of Use [Q2, $\chi2(2)$ =171.67, $p<0.001$], and Appropriateness for Application [Q3, $\chi2(2)$ =172.88, $p<0.001$].

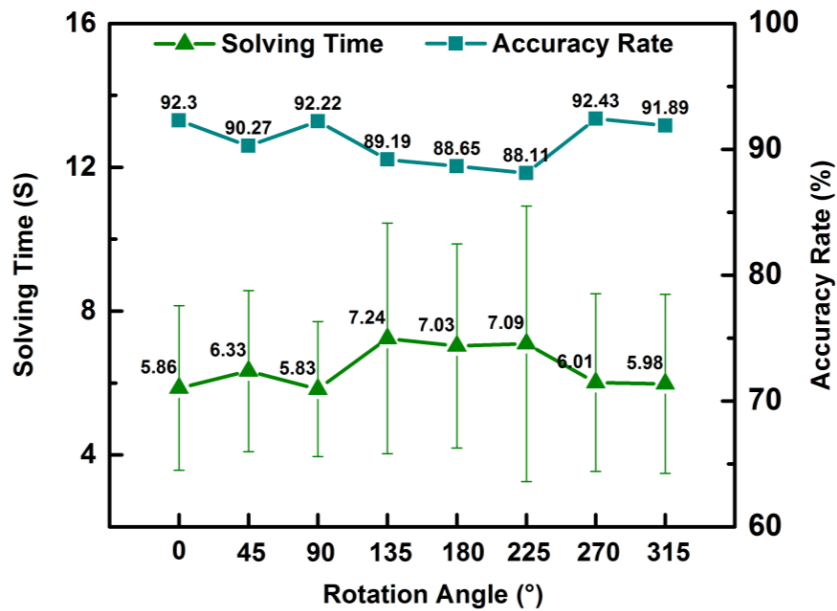

**Fig. 12**. The average solving time and accuracy rate CAPTCHAs based on images with different-angle contents: 0 °(reference group), 45 °, 90 °, 135 °, 180 °, 225 °, 270 °, and 315 °. Images were rotated clockwise.

**Table 7.** Participants' satisfaction of CAPTCHAs with contents of different angles

| | 0° | | 45° | | 90° | | 135° | |
|---|---|---|---|---|---|---|---|---|
| | AVG (0-5) | SD | AVG (0-5) | SD | AVG (0-5) | SD | AVG (0-5) | SD |
| **Q1.** It's visually comfortable | 4.63 | 0.85 | 3.50 | 0.94 | 2.43 | 1.07 | 1.97 | 0.96 |
| **Q2.** It's easy and efficient to recognize the images | 4.70 | 0.70 | 3.27 | 1.11 | 2.43 | 1.19 | 2.00 | 1.11 |
| **Q3.** It's appropriate for wide application | 4.57 | 0.82 | 3.27 | 1.05 | 2.27 | 1.08 | 1.87 | 1.01 |
| | 180° | | 225° | | 270° | | 315° | |
| | AVG (0-5) | SD | AVG (0-5) | SD | AVG (0-5) | SD | AVG (0-5) | SD |
| **Q1.** It's visually comfortable | 1.77 | 0.97 | 1.77 | 0.94 | 2.33 | 1.03 | 2.97 | 1.13 |
| **Q2.** It's easy and efficient to recognize the images | 1.87 | 1.11 | 1.77 | 1.04 | 2.37 | 1.16 | 3.03 | 1.10 |
| **Q3.** It's appropriate for wide application | 1.63 | 1.00 | 1.67 | 0.99 | 2.03 | 1.01 | 2.77 | 1.25 |

Generally, the tilting angle has no pronounced effects on the solving time and accuracy rate of Image CAPTCHAs. However, it is also interesting to note that, the solving time and accuracy rate for tilting angles of 135°, 180° and 225° are slightly degraded than all the other tilting angles, respectively. Tarr's cognitive theory [29] suggests that, during the recognizing of a rotated object, the visual input will be compared to a memorized counterpart that is closest to that rotation angle. The similar solving times of images that are titled within ±90° indicates that participants are equally familiar with the dogs and cats that are tilted to such angles. Therefore, the range of tilting angles is recommended to be within ±90° and larger tilting angles could degrade the efficiency and effectiveness of CAPTCHAs. However, satisfaction questionnaires show an even narrower range of tilting angle, ±45° and the most preferred tilting angle is 0°. All other tilting angles are rated as uncomfortable and less efficient in actual applications.

### 4.2.5 Image color

The averaged solving time (M) and its corresponding standard deviation (SD) are depicted in Fig. 13 for colored and monochrome images used in a CAPTCHA: colored images (M=5.86s, SD=2.29s) and monochrome images (M=6.02s, SD=2.27s). A repeated measures ANOVA with a Greenhouse-Geisser correction indicated that the two solving times are statistically different [F(1.00, 184.00)=6.41, $p$=0.012]. Fig. 13 also depicted the accuracy rate: colored images (92.30%), monochrome images (91.89%). The satisfaction questionnaires regarding different image colors are summarized in Table 8 in terms of (Q1) Visual Comfort, (Q2) Ease of Use and (Q3) Appropriateness for Application. Friedman tests return that the three dependent variables are all statistically significant: visual comfort [Q1, $\chi2(2)$ =27.00, $p$<0.001], ease of use [Q2, $\chi2(2)$ =29.00, $p$<0.001], and appropriateness for application [Q3, $\chi2(2)$ =28.00, $p$<0.001].

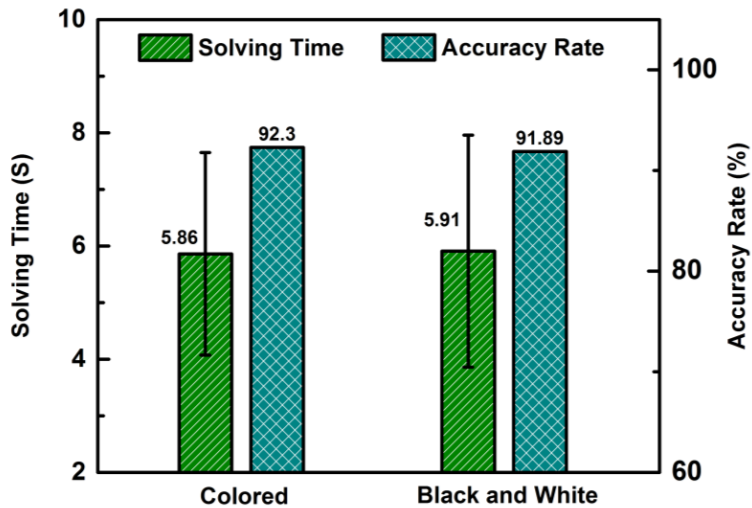

**Fig. 13.** Averaged solving time and accuracy rate CAPTCHAs consisted of colored images and black-and-white images.

**Table 8**. Satisfaction of CAPTCHAs consisted of different-style images.

|  | Colored | | Monochrome | |
| --- | --- | --- | --- | --- |
|  | AVG (0-5) | SD | AVG (0-5) | SD |
| **Q1.** It's visually comfortable | 4.30 | 0.95 | 2.73 | 1.42 |
| **Q2.** It's easy and efficient to recognize the mages | 4.37 | 0.89 | 2.70 | 1.02 |
| **Q3.** It's appropriate for wide application | 4.33 | 0.84 | 2.53 | 1.20 |

The similar solving time and accuracy rate of colored and monochrome images suggest that participants performed equally well for CAPTCHAs using both kinds of images. However, the satisfaction questionnaires show that monochrome images are not preferred. During the fact-to-face interview, more than 30% of the participants mentioned that, for monochrome images, dogs or cats with black or white furs are very likely to merge with the backgrounds and cause difficulty to identify them correctly. This may explain why monochrome images are rated with a low score of ~2.7 for all three variables.

**4.3 Summary for the design of usable image CAPTCHAs**

The design factors evaluated in terms of efficiency, effectiveness, satisfaction and eye-tracking are summarized in Table 9. For each evaluating dimension, the design factors are rated by different numbers of stars: "☆" indicates the parameter that is less preferred, "☆☆" means neutral and "☆☆☆" denotes a preferred parameter. In addition, "N/A" marked a comparison that is unavailable or the data are statistically insufficient to support a conclusion. It is worthy to note that each comparison is made within the same design factor.

**Table 9:** Image CAPTCHA design factors evaluated through efficiency, effectiveness, satisfaction and eye-tracking. "☆" indicates not recommended; "☆☆" means neutral; "☆☆☆" indicates it is most preferred and "N/A" marked the comparison is unavailable or the data is not statically significant to support a conclusion.

| Design Factors | | Efficiency | Effectiveness | Satisfaction | Eye-tracking |
|---|---|---|---|---|---|
| **Image Layout** | Square | N/A | ☆☆☆ | ☆☆☆ | ☆☆☆ |
| | Horizontal Rectangle | N/A | ☆☆☆ | ☆☆☆ | ☆☆☆ |
| | Vertical Rectangle | N/A | ☆☆☆ | ☆☆ | ☆☆☆ |
| | Upright Triangle | N/A | ☆☆☆ | ☆☆ | ☆☆ |
| | Inverted Triangle | N/A | ☆☆☆ | ☆ | ☆☆ |
| | Trapezoid | N/A | ☆ | ☆ | ☆ |
| **Number of Image** | 4 images | ☆☆☆ | ☆☆☆ | ☆☆☆ | ☆☆☆ |
| | 9 images | ☆☆☆ | ☆☆☆ | ☆☆☆ | ☆☆☆ |
| | 16 images | ☆☆ | ☆☆☆ | ☆☆ | ☆☆ |
| | 25 images | ☆ | ☆ | ☆ | ☆ |
| **Image Size** | 25 pixels × 25 pixels | ☆ | ☆ | ☆ | N/A |
| | 40 pixels × 40 pixels | ☆☆ | ☆☆☆ | ☆☆ | N/A |
| | 55 pixels × 55 pixels | ☆☆☆ | ☆☆☆ | ☆☆ | N/A |
| | 70 pixels × 70 pixels | ☆☆☆ | ☆☆☆ | ☆☆☆ | N/A |
| | 85 pixels × 85 pixels | ☆☆☆ | ☆☆☆ | ☆☆☆ | N/A |
| **Tilting of Image** | 0° | ☆☆☆ | ☆☆☆ | ☆☆☆ | N/A |
| | 45° | ☆☆☆ | ☆☆☆ | ☆☆☆ | N/A |
| | 90° | ☆☆☆ | ☆☆☆ | ☆ | N/A |
| | 135° | ☆☆ | ☆☆ | ☆ | N/A |
| | 180° | ☆☆ | ☆☆ | ☆ | N/A |
| | 225° | ☆☆ | ☆☆ | ☆ | N/A |
| | 270° | ☆☆☆ | ☆☆☆ | ☆ | N/A |
| | 315° | ☆☆☆ | ☆☆☆ | ☆☆ | N/A |
| **Image Color** | Colored | ☆☆☆ | ☆☆☆ | ☆☆☆ | N/A |
| | Monochrome | ☆☆☆ | ☆☆☆ | ☆ | N/A |

## 5. Conclusion

A four-dimensional usability evaluation method that measures eye-tracking, efficiency, effectiveness and satisfaction was proposed to investigate the universal design factors of Image CAPCHAs, including image layout, quantity, size, tilting angle and color. In the experiments, 37 participants were recruited to evaluate the five universal design factors and each design factor includes 2~8 variables. The cognitive processes revealed by eye-tracking indicate that participants are essentially insensitive to the variation of image contents and their attentions are also evenly distributed for different layouts. The average gazing time on each image decreased from 0.69 seconds to 0.40 seconds as the quantity of candidate images increased from 4 to 25, suggesting that users are less patient with more images. In addition, the gazing plot suggests that more than 70% of the participants reviewed CAPTCHA images row by row and less than 30% of them search randomly. It also turns out that reviewing candidate images with a particular sequence is more efficient than scanning randomly. The combination of eye-tracking results with efficiency, effectiveness and satisfaction tests essentially indicate that, square and horizontal rectangle are the preferred layouts; the quantity of candidate images may not exceed 16; the image size and tilting angle are suggested to be larger than 55 pixels $\times$ 55 pixels and within $\pm 45°$, respectively. Although colored and monochrome images are essentially the same in terms of efficiency and effectiveness, colored ones are more preferred because of its higher satisfaction. Those usability experiment results may serve as a design guideline that is expected to be helpful for developing usable image CAPTCHAs.

## Limitation of the study

Besides the variables studied in this paper, there are other factors that may influence the usability of Image CAPTCHAs, such as image resolution, transparency, noises like random dots and interfering curves, etc., also needed to be investigated. On the other hand, all participants were young, results presented here, therefore, may be kind of biased. To get more general results, a bigger and more representative pool of participants would be expected. Also, this study was conducted in a laboratory setting which is different from real-life scenes, the results here might have been kind of biased, too.

**Acknowledgements**. Junnan Yu gratefully thank Dr. Runze Li for helpful discussion. This work was supported by Shanghai Pujiang Program under Grant No. 13PJC072, Shanghai Philosophy and Social Science Program under Grant No. 2012BCK001, and Shanghai Jiao Tong University Interdisciplinary among Humanity, Social Science and Natural Science Fund under Grant No. 13JCY02.